\documentclass[
    ,final            
  ]
  {aipproc}

\layoutstyle{6x9}

\def\lsim{\raisebox{-.4ex}{$\stackrel{<}{\scriptstyle \sim}$\,}}

\usepackage{latexsym,epsfig}
\begin{document}
\input epsf
\title{Avoiding the dark energy coincidence problem with a
cosmic vector}


\author{Jose Beltr\'an Jim\'enez}{
  address={Departamento de  F\'{\i}sica Te\'orica,
 Universidad Complutense de
  Madrid, 28040 Madrid, Spain}
}

\author{Antonio L. Maroto}{
  address={Departamento de  F\'{\i}sica Te\'orica,
 Universidad Complutense de
  Madrid, 28040 Madrid, Spain}
}

\begin{abstract}
We show that vector theories on cosmological scales are excellent candidates for 
dark energy. We consider two different examples, both are theories 
with no dimensional parameters
nor potential terms, with natural initial conditions in the early universe
and the same number of free parameters as $\Lambda$CDM.
The first one exhibits  scaling behaviour during radiation and
a strong phantom phase today, ending in a "big-freeze" singularity.
This model provides the best fit to date for the SNIa Gold dataset.
The second theory we consider is standard electromagnetism. We show that a
 temporal  
electromagnetic field on cosmological scales 
generates an effective cosmological constant and that 
primordial electromagnetic quantum fluctuations 
produced during electroweak scale inflation could naturally explain, 
not only 
the presence of this field, but also the measured value of the dark 
energy density. The theory is compatible with all the local gravity tests, 
and is free from classical or quantum
instabilities. Thus,  not only the true nature of dark energy
could be established without resorting to new physics,  
but also the value of the cosmological
constant would find a natural explanation in the context of 
standard inflationary
cosmology.

\end{abstract}

\maketitle


\section{Introduction}
The fact that 
today matter and 
dark energy have comparable contributions to the energy density,  
$\rho_\Lambda\sim\rho_M\sim (2 \times 10^{-3}$ eV)$^4$ in natural units, 
poses one of the most important problems for models of dark energy. Thus,
if dark energy is a cosmological constant, its energy density
would remain  constant throughout the history 
of the universe,
whereas those of the rest of components (matter and radiation) grow as we 
go back in time.
Then the question arises as to whether it is a {\it coincidence} (or not) that 
they have comparable values today when they have differed 
by many orders of magnitude in  the past.  
Notice also that if $\Lambda$ is a fundamental constant of nature, its scale 
(around $10^{-3}$ eV) is more than 30 orders of magnitude smaller than 
the natural scale of  gravitation, $G=M_P^{-2}$ with $M_P\sim 10^{19}$ GeV. 
On the other hand, alternative models in which dark energy is a 
dynamical component rather than a 
cosmological constant also require the introduction of unnatural
scales in their Lagrangians or initial conditions in order
to account for the present phase of accelerated expansion. Such models 
are usually based on new physics, either in the form of 
new  cosmological fields or 
 modifications of Einstein's gravity 
\cite{quintessence1,quintessence2,quintessence3,quintessence4,
quintessence5} and they are generically plagued by additional problems
such as classical or quantum instabilities,
or inconsistencies with local gravity constraints.

Therefore, we would like to find a description for dark energy without  
dimensional scales 
(apart from Newton's constant $G$), with the same number of free parameters as 
$\Lambda$CDM, with natural initial conditions, with good fits to observations
and no consistency problems. In this work we will show that vector theories 
can do the job. With that purpose, we present two examples of such theories which
have been recently proposed \cite{BM,BMEM} (for other vector models 
see references in those works).

\section{Scaling vector dark energy}
The action in this case reads \cite{BM}:
\begin{eqnarray}
S=\int d^4x \sqrt{-g}\left(-\frac{R}{16\pi G}
-\frac{1}{4}F_{\mu\nu}F^{\mu\nu}-\frac{1}{2}(\nabla_\mu A^\mu)^2
+ R_{\mu\nu}A^\mu A^\nu\right)\label{scaling}
\label{action}
\end{eqnarray}
Notice that the theory contains no free parameters, the only dimensional
scale being the Newton's constant.   
The numerical factor
in front of the vector kinetic terms can be fixed 
by the field normalization.
Also notice that  the "mass" term 
 $R_{\mu\nu}A^\mu A^\nu$ can be written as a combination of derivative terms as 
$\nabla_\mu A^\mu\nabla_\nu A^\nu-\nabla_\mu A^\nu\nabla_\nu A^\mu$ and therefore
the theory contains no potential terms. This action resembles that of 
Maxwell electromagnetism in the Feynman gauge with a 
mass term.

The classical equations of motion derived from the action in (\ref{scaling})
are the Einstein's and
 vector field equations:
\begin{eqnarray}
R_{\mu\nu}-\frac{1}{2}R g_{\mu\nu}&=&8\pi G (T_{\mu\nu}+T_{\mu\nu}^A) \label{eqE}\\
\Box A_\mu + R_{\mu\nu}A^\nu&=&0 \label{eqA}
\end{eqnarray}
where $T_{\mu\nu}$ is the conserved energy-momentum tensor for matter and radiation and
$T_{\mu\nu}^A$ is the energy-momentum tensor coming from the vector field.
For the simplest isotropic and
 homogeneous flat cosmologies,  we assume that 
the spatial components of the vector field vanish, so that 
$A_\mu=(A_0(t),0,0,0)$ and that the 
space-time geometry will 
be given by:
\begin{equation}
ds^2=dt^2-a^2(t)\delta_{ij}dx^idx^j,
\label{metric}
\end{equation}
For this metric (\ref{eqA}) reads:
\begin{eqnarray}
\ddot{A}_0+3H\dot{A}_0-3\left[2H^2+\dot{H}\right]A_0=0\label{fieldeq0}
\end{eqnarray}
Assuming that the universe has gone through  radiation and 
matter phases in which
the contribution from dark energy was negligible, we can easily solve
this equation in those periods.  In that case, the
above equation has a growing and a decaying solution:
\begin{eqnarray}
A_0(t)=A_0^+t^{\alpha_+}+A_0^-t^{\alpha_-}\label{fieldsol}
\end{eqnarray}
with $A_{0}^\pm$ constants of integration and $\alpha_{\pm}=-(1\pm 1)/4$
 in the radiation era, and
 $\alpha_{\pm}=(-3\pm\sqrt{33})/6$ in the matter era. 
On the other hand, the $(00)$ component of Einstein's equations reads:
\begin{eqnarray}
H^2=\frac{8\pi G}{3}
\left[\sum_{\alpha=M,R} \rho_\alpha+\rho_{A}\right]
\label{Friedmann}\end{eqnarray}
where the vector energy density is given by: 
\begin{eqnarray}
\rho_{A}=\frac{3}{2}H^2A_0^2+3HA_0\dot A_0-\frac{1}{2}\dot A_0^2
\end{eqnarray}

\begin{figure}[h]
{\epsfxsize=6.5 cm\epsfbox{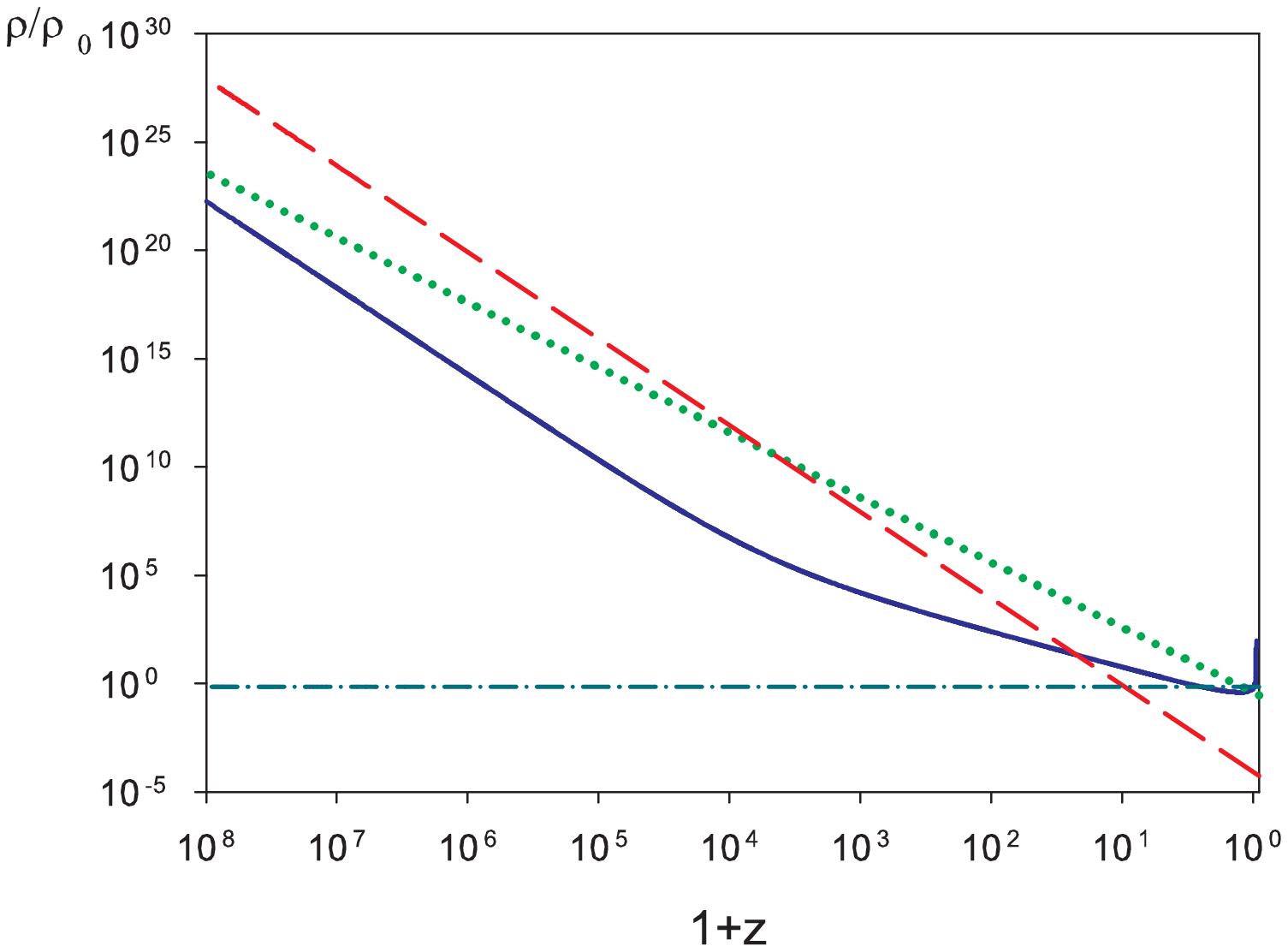}}\hspace{1.0cm}
{\epsfxsize=6.0 cm\epsfbox{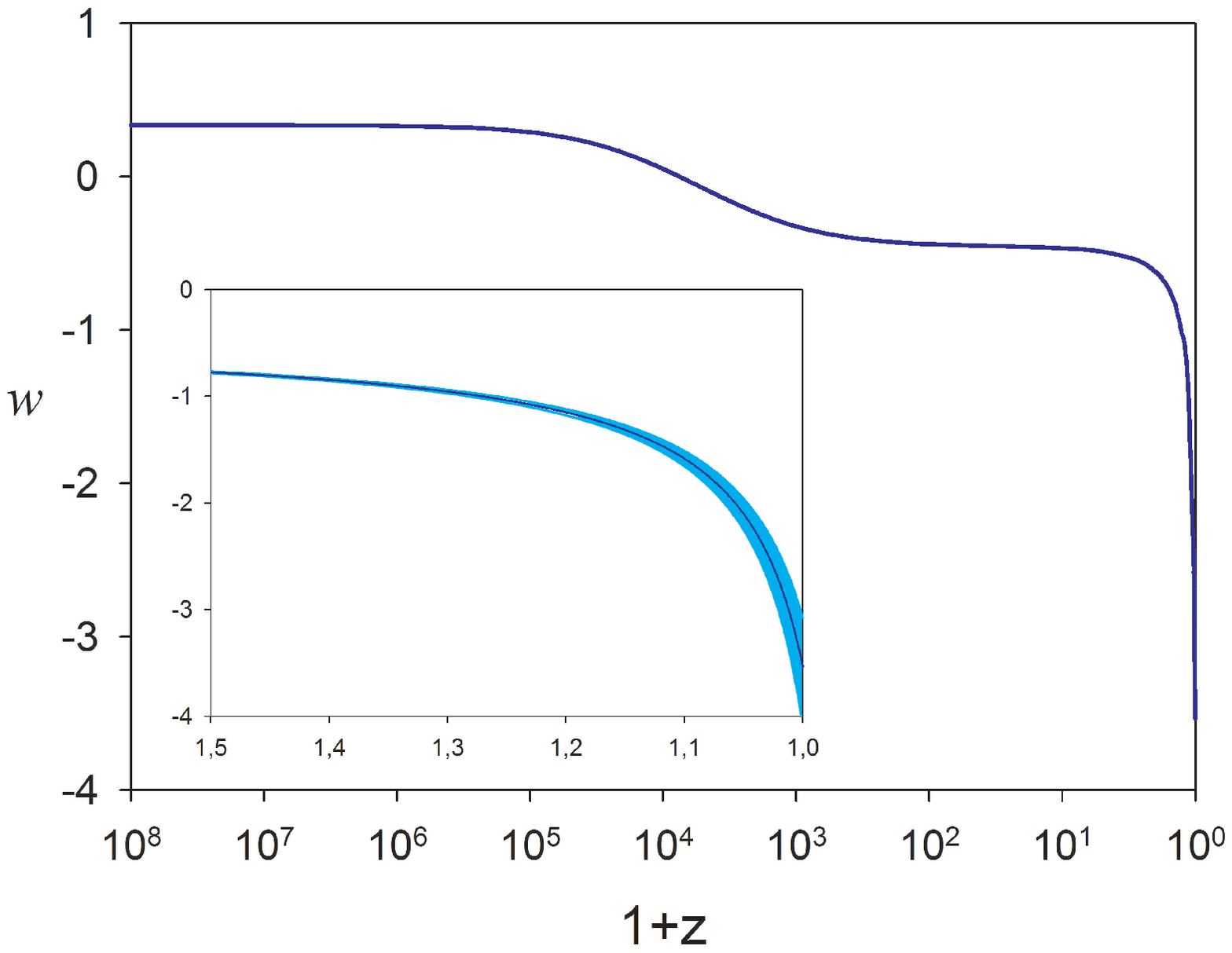}}
\end{figure}
{\footnotesize {\bf Figure 1:} (Left) Evolution of energy densities for the best fit model. 
Dashed (red) for
radiation, dotted (green) for matter and solid (blue) for vector 
dark energy. We show also for comparison the cosmological constant
density in dashed-dotted line.
(Right) Evolution of dark energy equation of state for
the best fit model. The lower panel shows the 1$\sigma$ confidence interval.}
\vspace{0.5cm}

Using the
growing mode solution from (\ref{fieldsol}), we obtain 
$\rho_{A}= \rho_{A0} a^\kappa$ 
with $\kappa=-4$ in the radiation era and 
$\kappa=(\sqrt{33}-9)/2 \simeq -1.63$ in 
the matter era. Thus, the energy density of the vector field 
starts scaling as radiation
 at early times, so that $\rho_A/\rho_R=$ const.  
However, when the universe enters its matter era, $\rho_A$ starts growing 
relative to 
$\rho_M$ eventually overcoming it at some point, in which the dark energy 
vector  
field would become
the dominant component (see Fig. 1).  Notice that  
since $A_0$ is essentially constant during 
radiation era, solutions do not depend on the precise initial time
at which we specify it.  Thus, 
once the present value of the Hubble parameter $H_0$
and the constant $A_0$ during radiation (which fixes the total matter
density $\Omega_M$) are specified, 
the model is completely 
determined, i.e. this model contains the same
number of parameters as $\Lambda$CDM, which is the minimum number 
of parameters
of a cosmological model with dark energy. As seen from
Fig.1  the evolution of the universe ends at a finite time $t_{end}$ where 
$a\rightarrow a_{end}$ with $a_{end}$ finite, 
$A_0(t_{end})=M_P/(4\sqrt{\pi})$,  
$\rho_{DE}\rightarrow \infty$ and
$p_{DE}\rightarrow -\infty$. This 
corresponds to a Type III ({\it big-freeze}) 
singularity according to the classification
in \cite{Nojiri}. 
 
We can also calculate the effective equation of state for dark energy
as:
\begin{eqnarray}
w_{DE}=\frac{p_A}{\rho_A}=\frac{-3\left(\frac{5}{2}H^2+\frac{4}{3}\dot{H}\right)A_0^2+HA_0\dot{A}_0
-\frac{3}{2}\dot{A}_0^2}{\frac{3}{2}H^2A_0^2+3HA_0\dot A_0-\frac{1}{2}\dot A_0^2}
\end{eqnarray}
Again, using the approximate solutions in (\ref{fieldsol}), we obtain:
$w_{DE}=1/3$ in the radiation era and $w_{DE}\simeq -0.457$ in the 
matter era. As shown in Fig. 1, the equation of state can cross the so called phantom 
divide, so that we can have $w_{DE}(z=0)<-1$.

In order to confront the  predictions of the model with observations
of high-redshift supernovae type Ia, we have carried out  a
$\chi^2$ statistical analysis for two  supernovae datasets, namely, 
the Gold set \cite{Gold}, 
containing 157 points with $z < 1.7$, and  
the more recent SNLS  data set \cite{SNLS}, comprising 115 
supernovae but with lower redshifts ($z < 1$). 

\begin{table}[h]
\vspace{0.5cm}
{\footnotesize\begin{tabular}{|c|c|c|c|c|}
\hline
  & & & &\\ 
 & VCDM  & $\Lambda$CDM  & VCDM & $\Lambda$CDM \\
& {Gold}  &{Gold} & 
{SNLS}   & {SNLS} \\
& & & &\\
\hline & & & & \\$\Omega_M$ &$0.388^{+0.023}_{-0.024}$ & $0.309^{+0.039}_{-0.037}$ & 
$0.388^{+0.022}_{-0.020}$ & $0.263^{+0.038}_{-0.036}$
\\& & && \\ 
\hline & & & &\\ $w_0$ &$-3.53^{+0.46}_{-0.57}$& $-1$ & $-3.53^{+0.44}_{-0.48}$ & $-1$
\\& & & & \\
\hline  & & & &\\$A_0$ & $3.71^{+0.022}_{-0.026}$ &--- & $3.71^{+0.020}_{-0.024}$&---
\\{$(10^{-4}\;M_P)$}& & & & \\
\hline  & & & &\\$z_T$ &$0.265^{+0.011}_{-0.012}$ &$0.648^{+0.101}_{-0.095}$ & $0.265^{+0.010}_{-0.012}$&$0.776^{+0.120}_{-0.108}$ 
\\& & & & \\
\hline  & & & &\\$t_0$   &$0.926^{+0.026}_{-0.023}$ & $0.956^{+0.035}_{-0.032}$& 
$0.926^{+0.022}_{-0.022}$ &$1.000^{+0.041}_{-0.037}$ 
\\ {$(H_0^{-1})$}& &  & & \\
\hline & & & &\\$t_{end}$ &$0.976^{+0.018}_{-0.014}$ & --- &$0.976^{+0.015}_{-0.013}$ & ---
 \\ {$(H_0^{-1})$}& & & &\\
\hline & & & &\\ $\chi^2_{min}$ & 172.9& 177.1 &115.8 & 111.0
\\ & & & &\\ 
\hline
\end{tabular}}
\end{table}

{\footnotesize {\bf Table 1:} Best fit parameters with $1\sigma$ 
intervals for the vector model (VCDM) and
the cosmological constant model ($\Lambda$CDM) for the Gold (157 SNe)
and SNLS (115 SNe) data sets. $w_0$
denotes the present equation of state of dark energy. $A_0$ is 
the constant value of the vector field component during radiation. 
$z_T$ is the deceleration-aceleration transition redshift.
$t_0$ is the age of the universe in units of the present
Hubble time. $t_{end}$ is the duration of the universe in 
the same units.}


\vspace{0.5cm}

In Table 1 we 
show the results for the best fit together 
with its corresponding $1\sigma$  intervals for the 
two data sets. We also
show for comparison the results for a standard
$\Lambda$CDM model. We see that the vector model (VCDM) fits the data
considerably
better than  $\Lambda$CDM (in more than $2\sigma$)  in the Gold 
set, whereas the situation is reversed in the SNLS set. This is just 
a reflection of the well-known $2\sigma$ tension \cite{tension}
between the two 
data sets.  Compared with $\Lambda$CDM, we see that VCDM
favors a younger universe (in $H_0^{-1}$ units)
with larger matter density. In addition, 
the deceleration-acceleration transition takes place at a lower redshift
in the VCDM case. The present value
of the equation of state  with $w_0=-3.53^{+0.46}_{-0.57}$ 
which clearly excludes the cosmological constant value $-1$. Future
surveys \cite{Trotta} are expected to be able to measure $w_0$ at the 
few percent level
and therefore could discriminate between the two models. 

We have also compared with other parametrizations
for the dark energy equation of state \cite{models}. Since our 
one-parameter fit has a reduced chi-squared: 
$\chi^2/d.o.f=1.108$,  VCDM provides 
the best fit to date for the Gold data set.

We see that unlike the 
cosmological constant case, throughout radiation era
 $\rho_{DE}/\rho_R\sim 10^{-6}$ in our case. Moreover the scale of the vector
field $A_0=3.71 \times 10^{-4}$ $M_P$ in that era is 
relatively close to the Planck scale and could arise naturally  
in the early universe without the need of introducing extremely small 
parameters (for instance in an inflationary epoch), 
thus avoiding the coincidence problem.

In order to study the model stability we have considered the evolution 
of metric and vector field perturbations. Thus, we obtain the dispersion
relation and  the propagation speed of  scalar, vector and tensor modes.  
For all of them we obtain $v=(1-16\pi G A_0^2)^{-1/2}$ which is
 real throughout the universe evolution, since the value $A_0^2=(16\pi G)^{-1}$
exactly corresponds  to that at the  final singularity. Therefore
 the model does not exhibit exponential instabilities. As shown in
\cite{Mukhanov}, the fact that the propagation speed is faster than $c$ 
does not necessarily implies inconsistencies with causality. We have 
also considered the evolution of scalar perturbations in the vector field 
generated by scalar metric perturbations during matter and radiation eras, 
and, again, we do not find exponentially 
growing modes.

If we are interested in extending the applicability range of the model
down to solar system scales then we should study the 
corresponding post-Newtonian parameters (PPN). We can see that
 for the model in (\ref{action}), the static PPN parameters
 agree with those of 
General Relativity \cite{Will}, i.e. 
$\gamma=\beta=1$. 
For the parameters  
associated to preferred frame effects we get: $\alpha_1=0$ and
$\alpha_2=8\pi  A_\odot^2/M_P^2$ where $A_\odot^2$ is the norm of the 
vector field at the solar system scale. Current limits 
$\alpha_2 \;\lsim 10^{-4}$ (or $\alpha_2\;\lsim 10^{-7}$ for static vector 
fields during
solar system formation) then impose a bound $A_\odot^2\lsim \, 
10^{-5}(10^{-8})
\,M_P^2$. In order to determine whether such bounds conflict with
the model predictions or not,  we should know the predicted value of 
the field at solar system scales, which in principle does not need to 
agree with the
cosmological value. 
Indeed, $A_\odot^2$  will be determined by the mechanism
that generated this field in the early universe characterized by its 
primordial spectrum of perturbations, and the
subsequent evolution in the formation of the galaxy and solar system. 
 Another potential difficulty arising generically in
 vector-tensor models is the presence of negative energy modes for
perturbations 
on sub-Hubble scales. They are known to lead to instabilities at the 
quantum level,  but not necessarily at the classical level as we have
shown previously. Work is in progress in order to determine which are the
 necessary conditions  to avoid the presence of such states in
this model.

\section{Is the nature of dark energy electromagnetic?} 
 
 In the previous section, we have seen that a generic vector field whose
action resembles that of electromagnetism with a mass term could be a good 
candidate for dark energy, but what about standard electromagnetism?. 
In a very recent work \cite{BMEM}, it has been shown using the 
covariant (Gupta-Bleuler)
formalism that it is indeed possible to explain cosmic
acceleration from the standard Maxwell's theory. 

We start by writing the standard electromagnetic action including a
gauge-fixing term in the presence of gravity:
\begin{eqnarray}
S&=&\int d^4x\sqrt{-g}\left[-\frac{1}{16\pi G}R-\frac{1}{4}F_{\mu
\nu}F^{\mu\nu}+\frac{\lambda}{2}\left(\nabla_\mu
A^\mu\right)^2\right]
\label{action}
\end{eqnarray}
The gauge-fixing term is required in order  
to define a consistent quantum theory for the electromagnetic
field \cite{Itzykson}, and we will
see that it plays a fundamental role on large scales. 
Still this action preserves a residual gauge symmetry 
$A_\mu\rightarrow A_\mu+\partial_\mu \phi$ with $\Box \phi=0$. 
Electromagnetic equations derived from this 
action can be written as:
\begin{eqnarray}
\nabla_\nu F^{\mu\nu}&+&\lambda\nabla^\mu\nabla_\nu
A^\nu=0\label{fieldeq}
\end{eqnarray}
 Notice that since 
we will be using the covariant Gupta-Bleuler formalism, we 
do not a priori impose the Lorentz condition. 

We shall first focus on the simplest case of a homogeneous electromagnetic 
field $A_\mu=(A_0(t),\vec A(t))$ in a flat Robertson-Walker background. 
In this space-time, equations (\ref{fieldeq}) read:
\begin{eqnarray}
\ddot{A}_0+3H\dot{A}_0+3\dot{H}A_0&=&0\nonumber\\
\ddot{\vec{A}}+H\dot{\vec{A}}&=&0\label{feqRW}
\end{eqnarray}  

We can solve (\ref{feqRW}) during the radiation and matter
dominated epochs when the Hubble parameter is given by $H=p/t$
with $p=1/2$ for radiation and $p=2/3$ for matter. In such a case
the solutions for (\ref{feqRW}) are:
\begin{eqnarray}
A_0(t)&=&A_0^+t+A_0^-t^{-3p}\label{A0sol}\\
\vec{A}(t)&=&\vec{A}^+t^{1-p}+\vec{A}^-\label{Azsol}
\end{eqnarray}
where $A_{0}^\pm$ and $\vec{A}^\pm$ are constants of integration.
Hence, the growing mode of the temporal component does not depend
on the epoch being always proportional to the cosmic time $t$, 
whereas the growing mode of the spatial component evolves as
$t^{1/2}$ during radiation and as $t^{1/3}$ during matter, i.e.
at late times the temporal component will dominate over the 
spatial ones.

The energy densities of the temporal and spatial components read:
\begin{eqnarray}
\rho_{A_0}&=&\lambda\left(\frac{9}{2}H^2A_0^2+3HA_0\dot{A}_0+\frac{1}{2}
\dot{A}_0^2\label{rhoA0}\right)
\\
\rho_{\vec{A}}&=&\frac{1}{2 a^2}(\dot{\vec{A}})^2
\end{eqnarray}
Notice that we need $\lambda>0$ in order to  
have positive energy density for $A_0$. In fact, it
is possible to show that imposing canonical normalization for the 
corresponding creation and annihilation operators we get $\lambda=1/3$
\cite{BMEM}. Besides, when inserting
the growing modes of the fields into these expressions we obtain
that $\rho_{A_0}=\rho_{A_0}^0$, $\rho_{\vec{A}}=\rho_{\vec{A}}^0\,a^{-4}$
and $\nabla_\mu A^\mu=$ const. 
Thus, the field behaves as a cosmological constant
throughout the evolution of the universe since its temporal
component gives rise to a constant energy density  whereas the
energy density corresponding to $\vec{A}$ always decays
as radiation. Moreover, this fact
 prevents the generation of a non-negligible anisotropy
which could spoil the highly isotropic  CMB radiation.
Finally, when the universe is dominated by
the electromagnetic field, both the Hubble parameter and 
$A_0$ become constant (one
can straightforwardly check that this is a solution of 
the complete system of equations) so the energy
density is also constant and the electromagnetic field behaves once again
as a cosmological constant leading therefore to a future de Sitter
universe. As the observed fraction of energy density associated to
a cosmological constant today is $\Omega_\Lambda\simeq 0.7$, we
obtain that the field value today must be $A_0(t_0)\simeq 0.3\,M_P$.

The effects of the high electric conductivity $\sigma$ can 
be introduced using the 
magneto-hydro\-dynamical approximation and including the current term
$J_i=
\sigma(\partial_0  A_i-\partial_i A_0)$ on the r.h.s. of  
Maxwell's equations. 
Notice that because of the universe electric 
 neutrality,  conductivity does
not affect the evolution of $A_0(t)$. The infinite conductivity limit
simply eliminates the growing mode of $\vec A(t)$ in (\ref{Azsol}).

We still need to understand which are the appropriate initial conditions 
leading to the  present value of $A_0$. In order to avoid the 
cosmic coincidence problem, such initial conditions 
 should have been set in a natural way in the 
early universe.  
In a very interesting work \cite{Arkani},
 it was suggested that the present value 
of the dark energy density could be related to physics at the 
electroweak scale since $\rho_\Lambda \sim (M_{EW}^2/M_P)^4$, where
$M_{EW}\sim 10^3$ GeV. This relation offers 
a hint 
on the possible mechanism generating the initial
amplitude of the electromagnetic fluctuations. Indeed, we see that if 
 such amplitude is set
by the size of the Hubble horizon at the electroweak era, i.e.
$A_0(t_{EW})^2\sim H_{EW}^2$, then
the correct scale for the dark energy density is obtained.
Thus, using the Friedmann equation, we find $H_{EW}^2\sim M_{EW}^4/M_P^2$, but
according to (\ref{rhoA0}), $\rho_{A_0}\sim H^2A_0^2\sim\;$const., 
so that $\rho_{A_0}\sim H_{EW}^4\sim (M_{EW}^2/M_P)^4$
as commented before. 

A possible implementation of this mechanism can take place 
during inflation. 
Notice that 
the typical
scale of the dispersion of quantum field fluctuations on super-Hubble 
scales generated 
in an inflationary period is precisely set
by the almost constant Hubble parameter during such period $H_I$, 
i.e. $\langle A_0^2\rangle\sim H_I^2$ \cite{Linde}.  The
 correct dark energy density can then be 
naturally obtained if initial conditions for the 
electromagnetic fluctuations are set during an  inflationary 
epoch at the scale $M_I\sim M_{EW}$.

Despite the fact that the background evolution in the present case 
is the same as
in $\Lambda$CDM, the evolution of metric perturbations could
be different, thus offering an observational way of discriminating
between the two models. In fact, the evolution of the scalar
perturbation $\Phi_k$
with respect to the $\Lambda$CDM model gives rise to 
a possible discriminating  contribution to the 
late-time integrated Sachs-Wolfe effect \cite{Turok}. The propagation speeds 
of scalar, vector and tensor perturbations are found
to be real and equal to the speed of light, so that the theory is 
classically stable. We have also checked that  the theory does
not contain ghosts and it is therefore stable at the quantum level.    
On the other hand, using the explicit expressions in \cite{Will} for 
the  vector-tensor theory of gravity corresponding to the action in 
(\ref{action}),  it is
possible to see that all the parametrized post-Newtonian (PPN) parameters
agree with those of General Relativity,  i.e. the theory is compatible
with all the local gravity constraints for any value 
of the homogeneous background vector field \cite{VT}. 

The presence of large scale electric fields generated by inhomogeneities
in the $A_0$ field opens also the possibility 
for the generation of large scale currents which in turn
could contribute to the presence of magnetic fields
with large 
coherence scales. This could shed light on the problem of explaining the 
origin of cosmological magnetic fields. 
Work is in progress in this direction.

\section{Conclusions}

We have shown that  vector theories offer a simple and accurate 
description of dark energy in which the coincidence problem
 could be easily avoided. In our first example, the scaling behaviour
during radiation and the natural initial conditions for the vector
field offer a neat way around the problem. Moreover, in our 
second example, the presence of a cosmological 
electromagnetic field generated during inflation provides a natural  
explanation for the cosmic acceleration. This result not 
only offers a solution to the problem of establishing the true nature of 
dark energy, but also explains the value of the cosmological constant 
without resorting to new physics. In this scenario the 
fact that matter and dark energy densities {\it coincide} today 
is just a consequence of inflation taking place at the electroweak scale. 
Present and
forthcoming astrophysical and cosmological observations will 
be able to discriminate these proposals from the standard $\Lambda$CDM
cosmology.  

\vspace{0.5cm}

{\em Acknowledgments:} This work has been  supported by
DGICYT (Spain) project numbers FPA 2004-02602 and FPA
2005-02327, UCM-Santander PR34/07-15875 and by CAM/UCM 910309. 
J.B. aknowledges support from MEC grant
BES-2006-12059.


\vspace{0.2cm}
\begin{thebibliography}{99}
\bibitem{quintessence1} C. Wetterich, {\it Nucl. Phys.} {\bf B302}, 668 (1988); 
\bibitem{quintessence2}R.R. Caldwell, R. Dave and P.J. Steinhardt, {\it Phys. Rev. Lett.}
 {\bf 80}, 1582 (1998)
\bibitem{quintessence3} C. Armendariz-Picon, T. Damour and V. Mukhanov, 
{\it Phys. Lett.} {\bf B458}, 209 (1999)
\bibitem{quintessence4} S.M. Carroll, V. Duvvuri, M. Trodden, M.S. Turner, 
{\it Phys. Rev.} {\bf D70}: 043528, (2004)
\bibitem{quintessence5} G. Dvali, G. Gabadadze and M. Porrati, 
{\it Phys. Lett.} {\bf B485}, 208 (2000)
\bibitem{BM} J. Beltr\'an Jim\'enez and A.L. Maroto, {\it Phys. Rev.} 
{\bf D78}, 063005 (2008)  and arXiv:0807.2528 [astro-ph]
\bibitem{BMEM} J. Beltr\'an Jim\'enez and A.L. Maroto, 
arXiv:0811.0566 [astro-ph]
\bibitem{Nojiri} S.~Nojiri, S.~D.~Odintsov and S.~Tsujikawa,
  {\it Phys.\ Rev.} {\bf D71} (2005) 063004; 
M.~Bouhmadi-L\'opez, P.~F.~Gonz\'alez-D\'{\i}az and P.~Mart\'{\i}n-Moruno,
  {\it Phys. Lett.} {\bf B659} (2008) 1
\bibitem{Gold} A.G. Riess at al. {\it Astrophys.J.} {\bf 607}, 665
(2004)
\bibitem{SNLS} P. Astier et al.,  {\it Astron. Astrophys.}
{\bf 447}: 31-48, (2006). 
\bibitem{tension} S. Nesseris and L. Perivolaropoulos, {\it JCAP} 
{\bf 0702}: 025, (2007). 
\bibitem{Trotta} R. Trotta and R. Bower, {\it Astron. Geophys.}
{\bf 47}: 4:20-4:27, (2006)
\bibitem{models}R. Lazkoz, S. Nesseris, L. Perivolaropoulos, {\it JCAP} 
{\bf 0511}:010, (2005) 
\bibitem{Mukhanov}
  E.~Babichev, V.~Mukhanov and A.~Vikman, JHEP {\bf 0802}, 101 (2008)
\bibitem{Will} C. Will, {\it Theory and experiment in gravitational physics},
Cambridge University Press, (1993)
\bibitem{Itzykson} C. Itzykson and J.B. Zuber, {\it Quantum Field Theory},  
McGraw-Hill (1980) 
\bibitem{Arkani}
  N.~Arkani-Hamed, L.~J.~Hall, C.~F.~Kolda and H.~Murayama,
{\it Phys.\ Rev.\ Lett.}\  {\bf 85} (2000) 4434
\bibitem{Linde} A. Linde, {\it Particle physics and inflationary
cosmology}, Harwood Academic Press (1996)
\bibitem{Turok} R.G. Crittenden and N. Turok, {\it Phys.\ Rev.\ Lett.}\  {\bf 76} 
(1996) 575
\bibitem{VT} J. Beltr\'an Jim\'enez and A.L. Maroto, 
arXiv:0811.0784 [astro-ph]










\end{thebibliography}
\end{document}